\documentclass[aps,pra,twocolumn,english,preprint,10pt,nofootinbib]{revtex4}
\usepackage[T1]{fontenc}
\usepackage[latin9]{inputenc}
\usepackage{babel}
\usepackage{nccmath}

\usepackage{graphicx,amsmath,amssymb,color}
\usepackage[normalem]{ulem}
\usepackage{amsfonts}
\usepackage[toc,page]{appendix}
\usepackage{hyperref}
\usepackage{latexsym}
\usepackage{amsfonts}
\usepackage{algpseudocode}
\usepackage{amsthm}
\usepackage{mathrsfs}
\usepackage{natbib}
\usepackage{color,verbatim}
\usepackage{psfrag}
\usepackage[svgnames]{xcolor}
\usepackage{tikz}

\usepackage[svgnames]{xcolor}
\usepackage{tikz}
\usetikzlibrary{decorations.markings}
\usetikzlibrary{shapes.geometric}

\pgfdeclarelayer{edgelayer}
\pgfdeclarelayer{nodelayer}
\pgfsetlayers{edgelayer,nodelayer,main}

\tikzstyle{none}=[inner sep=0pt]

\tikzstyle{no_field_qubit}=[fill=white, draw=black, shape=circle,minimum size=0.5cm]
\tikzstyle{plus_field_qubit}=[fill={rgb,255: red,255; green,128; blue,128}, draw=black,shape=circle,label=center:{\tiny \textcolor{black}{$-\lambda$}},minimum size=0.5cm]
\tikzstyle{minus_field_qubit}=[fill={rgb,255: red,128; green,128; blue,255}, draw=black, shape=circle,label=center:{\tiny \textcolor{black}{$+\lambda$}},minimum size=0.5cm]
\tikzstyle{plus_filed_nolabel}=[fill={rgb,255: red,255; green,128; blue,128}, draw=black,shape=circle,minimum size=0.5cm]
\tikzstyle{minus_field_nolabel}=[fill={rgb,255: red,128; green,128; blue,255}, draw=black, shape=circle,minimum size=0.5cm]
\tikzstyle{coloring_node}=[fill=white, draw=black, shape=circle,minimum size=1cm]
\tikzstyle{one_hot_node}=[fill={rgb,255: red,255; green,128; blue,255}, draw=black, shape=circle,minimum size=0.3cm]
\tikzstyle{dot_dot_dot}=[fill=none, draw=none, shape=rectangle, label=center:{\Huge \textcolor{black}{...}},minimum size=0.5cm]
\tikzstyle{edge_end}=[fill=none, draw=none, shape=circle]
\tikzstyle{virtual_plus}=[fill={rgb,255: red,255; green,128; blue,128}, draw=black, shape=circle,label=center:{\tiny \textcolor{black}{$-\infty$}},minimum size=0.5cm]
\tikzstyle{virtual_minus}=[fill={rgb,255: red,128; green,128; blue,255}, draw=black, shape=circle,label=center:{\tiny \textcolor{black}{$+\infty$}},minimum size=0.5cm]
\tikzstyle{dw_and_one_hot_qb}=[fill={rgb,255: red,0; green,0; blue,0}, draw=black, shape=circle]
\tikzstyle{one_hot_only_qb}=[fill={rgb,255: red,255; green,0; blue,255}, draw=black, shape=circle]
\tikzstyle{distance_node}=[fill=yellow, draw=black, shape=circle,minimum size=0.5cm]

\tikzstyle{label_1}=[fill=none, draw=none, shape=rectangle]
\tikzstyle{label_2}=[fill=none, draw=none, shape=rectangle]
\tikzstyle{label_3}=[fill=none, draw=none, shape=rectangle]
\tikzstyle{label_4}=[fill=none, draw=none, shape=rectangle]
\tikzstyle{label_5}=[fill=none, draw=none, shape=rectangle]
\tikzstyle{label_6}=[fill=none, draw=none, shape=rectangle]

\tikzstyle{ferro_coupling}=[-,draw=black]
\tikzstyle{anti_ferro_coup}=[-, draw=red]
\tikzstyle{coloring_edge}=[-, draw=black,line width=2]
\tikzstyle{one_hot_edge}=[-, draw={rgb,255: red,0; green,128; blue,0}]
\tikzstyle{domain_wall_edge}=[-,draw=black,line width=2]
\tikzstyle{dw_and_one_hot_edge}=[-, draw={rgb,255: red,0; green,0; blue,0}]
\tikzstyle{one_hot_only_edge}=[-, draw={rgb,255: red,255; green,0; blue,255}]

\newcommand{\ket}[1]{|#1 \rangle}

\preprint{IPPP/20/8}

\begin{document}

\title{Quantum Computing for Quantum Tunnelling}
\author{Steven Abel}
\email{s.a.abel@durham.ac.uk}
\address{Institute for Particle Physics Phenomenology, and Department of Mathematical Sciences, Durham University, Durham DH1 3LE, U.K.}

\author{Nicholas Chancellor}
\email{nicholas.chancellor@gmail.com}
\address{Department of Physics and Durham Newcastle Joint Quantum Centre\\ Durham University, South Road, Durham, UK}

\author{Michael Spannowsky}
\email{michael.spannowsky@durham.ac.uk}
\address{Institute for Particle Physics Phenomenology, Department of Physics, Durham University, Durham DH1 3LE, U.K.}

\begin{abstract}
We demonstrate how quantum field theory problems can be embedded on quantum annealers. The general method we use is a discretisation of the field theory problem into a general Ising model, with the continuous field values being encoded into Ising spin chains. To illustrate the method, and as a simple proof of principle, we use a (hybrid) quantum annealer to recover the correct profile of the thin-wall tunnelling solution. This method is applicable to many nonperturbative problems.
\end{abstract}

\maketitle


\section{Introduction}

There has been increasing interest in the possibility of simulating Quantum Field Theory (QFT) on quantum computers  \cite{Feynman:1981tf}, with the development 
of efficient algorithms to compute scattering probabilities in simple theories of scalars and fermions  \cite{Zalka:1996st,Jordan:2011ne,Jordan:2011ci,Garcia-Alvarez:2014uda,Jordan:2014tma,Beverland:2014gpa,Jordan:2017lea,Moosavian:2017tkv,Preskill:2018fag,Lamm:2018siq,Yeter-Aydeniz:2018mix,Bauer:2019qxa,Moosavian:2019rxg,Gustafson:2019vsd,Klco:2019yrb,Harmalkar:2020mpd}.
In particular it is known that by latticizing field theories, quantum computers should be able to compute scattering probabilities in QFTs with a run time that is polynomial in the desired precision, and in principle to a precision that is not bounded by the limits of perturbation theory. However a particularly difficult aspect of this programme is the preparation of scattering states
\cite{Jordan:2011ci,Jordan:2014tma,Garcia-Alvarez:2014uda,Jordan:2017lea,Moosavian:2017tkv,Moosavian:2019rxg,Gustafson:2019vsd,Klco:2019yrb,Harmalkar:2020mpd}, with several works having proposed a hybrid classical/quantum approach to solving this problem \cite{Lamm:2018siq,Harmalkar:2020mpd,Wei:2019rqy,Matchev:2020wwx}. A complementary approach is to map field theory equations to discrete quantum walks \cite{Arrighi:2018PRA,Marque-Martin:2018PRA,Jay:2019PRA,DiMolfetta:2020QIP} which can be simulated on a universal quantum computer.

In this paper we point out that 
certain nonperturbative quantum processes do not suffer from this difficulty, and lend themselves much more readily to study on quantum computers in the short term. These are the tunnelling and related processes, which are of fundamental importance for the explanation of quantum mechanical and quantum field theoretical phenomena, for example transmission rates of electron microsopes, first-order phase transitions during baryogenesis, or the potential initiation of stochastic gravitational wave spectra in the early Universe and many more.

Typically in tunnelling, the system begins in a false vacuum state that is non-dynamical and virtually trivial. The initial state can be very long lived, with tunnelling to a lower ``true'' vacuum state taking place via non-perturbative instanton configurations.
In principle in such a process, the confinement of the initial state to a false vacuum prepares the state for us, so that the 
analytically straighforward perturbative phenomena are paradoxically the quantum computationally more difficult ones.

As opposed to quantum computing realised by a series of discrete ``gate'' operations, quantum annealers \cite{farhi00a,neven2009nips} perform continuous time quantum computations, and therefore they are well-suited to the study of tunnelling problems by direct simulation (although our discussion could ultimately 
be adapted to gate-model quantum computers as well) \cite{finilla94a,kadowaki98a,brooke99a,dickson13a,lanting14a,albash15a,albash16a,boixo16a,chancellor16b,Benedetti16a,Muthukrishnan2016}.
In particular these devices, produced by D-Wave Systems \cite{LantingAQC2017}, can be seeded with initial conditions using the ``reverse annealing'' feature,\cite{chancellor17b} allowing the simulation of dynamics. In contrast with the quantum-gate devices, 
 they are already quite large, $2048$ qubits in the current generation, with work ongoing to 
develop much more connected  $5000$ qubit machines. Moreover they operate in a dissipative rather than fully coherent regime, which is likely to be realistic for many real theories in which there are interactions with matter. In the present context this would be relevant for studies of so-called {\it thermal} tunnelling rather than (or in addition to) quantum tunnelling. 
 D-Wave devices have been   able successfully to simulate condensed matter systems, sometimes showing advantages over classical counterparts \cite{Harris:2018Science,King:2018Nature,King:2019arXiv}.

The main objective of this work is to demonstrate how a field theory 
problem can be successfully encoded on a quantum annealing device, and to do this we will focus on the classic problem 
of obtaining tunnelling rates for a system stuck in a metastable minimum (a.k.a. false vacuum).

{\begin{figure}
  \includegraphics[width=4.2cm]{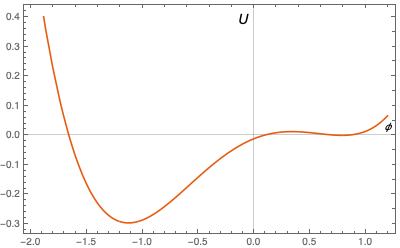}
  \includegraphics[width=4.2cm]{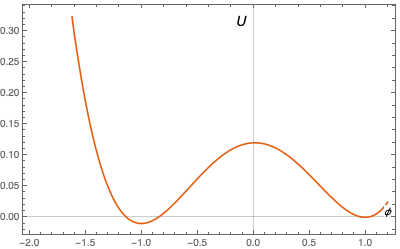}
  \caption{The thick-wall potential left (with $\epsilon = 0.3$, and true and false minima at $\phi_- = -1.12542$ and $\phi_+ = 0.786483$ respectively), and thin-wall potential right (with $\epsilon=0.01$).}
  \label{fig:potential}
\end{figure}
\begin{figure}
  \includegraphics[width=4.2cm]{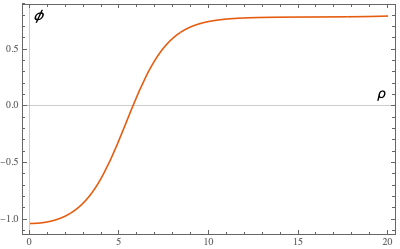}
  \includegraphics[width=4.2cm]{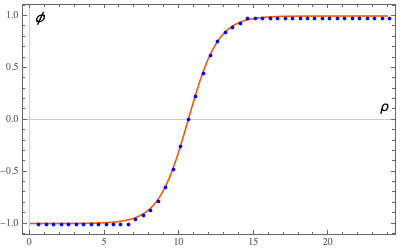}
  \caption{Solutions for the thick- and thin-wall potentials. The thin-wall solution computed using the hybrid quantum-classical techniques as discussed later is overlaid on the right panel.}
  \label{fig:solutions}
\end{figure}}

\section{Set-up of a simple problem}
\label{sec:problem}

A useful potential to focus on is the following quartic one:
\begin{equation}
\label{eq:pot4}
V(\phi)\,=\,\frac{\lambda}{8} (\phi^2 - a^2)^2 + \frac{\epsilon}{2a}(\phi -a)~.
\end{equation} 
The potential is shown in Fig.\ref{fig:potential}. On the left we show the ``thick-wall'' regime where $\epsilon$ is large. This limit is when the barrier is close to disappearing (or has disappeared altogether) and the walls become comparable in size to the bubble itself.  For numerics we choose $a=\lambda=1$ and $\epsilon=0.3$. The opposite ``thin-wall'' regime (for which we choose $\epsilon=0.01$) is the limit in which $\epsilon$ is small and is approximately the difference in vacuum energy density between the false and true minima.  

We are interested in the situation where the system starts in the false vacuum, and our objective is to study the rate  per unit volume  of tunnelling out of it.  
The analytic calculation of this rate is a classic problem, but it is worth briefly recapping it in order to recast the result in a form that can easily be compared with
the results from a quantum simulation. It proceeds as follows.

First let us remove the extraneous constant term by working with $U(\phi)  \,=\, V(\phi )\, -\, V(\phi_+),$
which has $U(\phi_+)=0$. Using the well-known technique of
 \cite{Coleman:1977py,Callan:1977pt,0521318270,Linde:1981zj}, the bubble profile 
 is given by finding a ``bounce solution'' to the following differential equation:
\begin{equation}
\frac{d^2 \phi}{d \rho^2} + \frac{c}{\rho} \frac{d\phi}{d\rho} \,=\,  U'~,
\label{eq:bubble}
\end{equation}
where in four dimensions, $c$ takes the value $2$ or $3$ for a finite temperature $O(3)$ symmetric bubble, or a purely quantum tunnelling $O(4)$ symmetric instanton, respectively. The 
required ``bounce'' is subject to the boundary condition that $d\phi/d\rho=0$  as $\rho\rightarrow 0,\infty$, which determines the starting value 
$\phi(0)$, which is the field-value at the centre of the radially symmetric bubble or instanton (also called the escape-point). The resulting $\phi(\rho)$ profile for our particular choice of parameters is shown in Fig.~\ref{fig:solutions}. 

Once such a solution is determined, the tunnelling rate per unit volume can be estimated from its classical action:
\begin{align}
\Gamma_4 & \,=\, A_4\, e^{-S_4[\phi]}\, ,\nonumber \\
\Gamma_3 & \,=\, A_3 \, T e^{-S_3[\phi]/T }\, ,
\end{align}
respectively. The quantum determinant prefactors $A_4,A_3$ are notoriously difficult to calculate, but for our purposes it will be sufficient to focus on the influence of the classical action. 

The expressions for the action can be expressed in simple analytic terms in the two limits. In the thick wall limit the bounce action can be accurately approximated  by expanding around the  value   $\epsilon=\epsilon_0$, above which the barrier disappears (i.e. when the discriminant vanishes), which gives a cubic potential about the false vacuum. 
This critical value corresponds to $\epsilon_{0}=2\lambda a^{4}/3\sqrt{3}$.
Defining $\rho = \sqrt{ {2}/{3} ( 1-\epsilon/\epsilon_0) } $, the location of the minima is 
\begin{align}
\frac{\phi_{+}}{a} & ~=~\frac{1+\rho}{\sqrt{3}} +\mathcal{O}(\rho^{2})\,,\nonumber \\
\frac{\phi_{-}}{a} & ~=~-\frac{2}{\sqrt{3}}+\mathcal{O}(\rho^{2})\,.
\end{align}
Then following the rescaling procedure of \cite{Linde:1981zj}, 
the tunnelling actions for the $O(4)$ and $O(3)$ symmetric
solutions can be written in terms of standard actions:
\begin{align}
S_{4} & =\frac{3\rho}{\lambda}S_{4}^{0}\,\,\,\,;\,\,\,\,\,\,S_{4}^{0}=91\nonumber \\
S_{3} & =\frac{3a\rho^{3/2}}{\lambda^{1/2}}S_{3}^{0}\,\,\,\,;\,\,\,\,\,\,S_{3}^{0}=19.4
\end{align}
The thin-wall regime is somewhat easier to study numerically, and semi-analytically the actions can be expressed in terms of the action $S_1$ for the one-dimensional $c=0$ problem~\footnote{This is also the energy of the physical ``domain wall'' solution, but for reasons that will become apparent it would be confusing to use this terminology.}:
\begin{align}
S_{4} & =\frac{27 \pi^2 S_1^4}{2 \epsilon^3} \,\,\,\,;\,\,\,\,\,
S_{3} =\frac{16 \pi^3 S_1^3}{3 \epsilon^2}~.
\end{align}

These limiting regimes give simple power-law behaviour for the tunnelling actions, against which the scaling of the (logarithm of) 
tunnelling rates could be tested, providing a useful laboratory for directly studying quantum annealing results.

As we stated in the introduction, the purpose if this study is not to recover these classical instanton solutions for the tunnelling {\it per se}, as they are well-known, but rather 
to demonstrate that the corresponding field-theory configuration can be suitably encoded into a quantum annealer. Once we have established this as a working principle, one could even envisage 
testing for the above behaviour directly. Therefore we will in what follows focus on using a quantum annealer to recover the simple $c=0$ solution required for the thin-wall regime, as a proof of principle.
We will therefore set ourselves the task of minimising the corresponding action integral,
\begin{equation}
S_1\,=\,\int_0^\infty d\rho ~\frac{1}{2} {\dot\phi^2}  + U(\phi)   ~,
\label{eq:bubble1}
\end{equation}
which should yield a solution of the form shown in Fig.\ref{fig:solutions}b.

\section{Encoding the field theory}
\label{sec:encode}

Let us start with the central problem, which is how to formulate a continuous scalar field theory on quantum annealers. 
A quantum annealer is based on the adiabatic theorem of quantum mechanics, which implies that a physical system will remain in the ground state if a given perturbation acts slowly enough, and if there is a gap between the ground state and the rest of the system's energy spectrum \cite{farhi00a}. For the annealer to provide a solution to a mathematical problem, e.g. the calculation of $\phi(\rho)$ for Eq.~\ref{eq:bubble1}, we have to find a mapping such that the expectation value of its Hamiltonian can be identified with its solution, i.e. that it allows in this example to identify 
\begin{equation}
\phi(\rho) \iff \lim_{t\to 0} \left <\mathcal{H}_{\mathrm{QA}}(t) \right >.
\label{eq:HAQphi}
\end{equation}

The form of the Hamiltonian available to a quantum annealer  is that of a general Ising model, in addition to a time-dependent transverse field:
\begin{equation}
\mathcal{H}_{\mathrm{QA}}(t) =  \sum_i \sum_j J_{ij} \sigma_i^Z \sigma_j^Z  + \sum_i h_i \sigma_i^Z + \Delta(t) \sum_{i} \sigma_i^X~,
\label{eq:Hquantan}
\end{equation}
where $\sigma^Z_i=\left(\begin{array}{cc} 1  & 0 \\ 0 & -1\end{array}\right)$ ($\sigma^Z \ket{0}=\ket{0}$,~$\sigma^Z \ket{1}=-\ket{1}$) is the Pauli $Z$ operator, with the subscript indicating which spin it acts upon, and $\sigma^X$ is its friend pointing in the $X$-direction.
The gradual decrease of $\Delta(t) \to 0$ from a large value should drive the system into the ground state of the time-independent 
part of the Hamiltonian, and this is where we will put the field theory:
\begin{equation}
\mathcal{H}=\sum_i\sum_j J_{ij} \sigma^Z_i\sigma^Z_j+\sum_i h_i\sigma^Z_i \label{eq:H_two_body}~.
\end{equation}
It is worth noting that the couplings $J_{ij}$ and $h_i$ could also be adiabatically adjusted in the annealing process, and this could ultimately be used to adjust the potential $U(\phi)$ of a system in the quantum annealer so as to observe tunnelling, assuming it can be encoded. We will further split the Hamiltonian into three generic pieces, as 
 \begin{equation}
\mathcal{H}~=~ \mathcal{H}^{(\mathrm{chain})}+ \mathcal{H}^{\mathrm{(QFT)}} + \mathcal{H}^{\mathrm{(BC)}}.
\label{eq:Hp}
\end{equation}
Here, $\mathcal{H}^{\mathrm{(QFT)}}$ is the Hamiltonian corresponding to the minimisation of the action in Eq.~\ref{eq:bubble1} and $\mathcal{H}^{\mathrm{(BC)}}$ is a Hamiltonian that we add to enforce the boundary conditions\footnote{For a classical neural network-based approach to solving Eq.~\ref{eq:bubble} by treating it as an optimisation problem see \cite{Piscopo:2019txs}.}. 

However our first task is to encode continuous field values over a 
continuous domain, with only the discrete Ising model to hand: this is what ${\mathcal{H}}^{(\mathrm{chain})}$ is for. 
We begin by splitting the radius variable $\rho$ into $M\gg 1$ discrete values and the field value at the $\ell$'th position into 
$N\gg 1$ discrete values:
\begin{equation}
\begin{array}{lll}
\rho_\ell & \,=\, \ell\nu ~~& =~~ \nu \ldots M \nu\nonumber \\
\phi (\rho_l) &\, =\, \phi_0 + \alpha_l \xi ~~& =~~ \phi_0+\xi \, \ldots \, \phi_0 + N \xi\, ,
\end{array}
\end{equation}
where in the present context one might for example take a fiducial value $\phi_0\approx  -a$ and $\xi=2a /N$, with $M \nu =\Delta \rho$. Thus our Ising interaction $J_{ij}$ 
is an $(MN)\times (MN)$ matrix, while $h_i$ is an $(NM)$-vector. 

We must now separate those spins in the annealer that correspond to fields at different values of $\ell$, effectively splitting $J_{ij}$ and $h_i$ into $N\times N$ sub-blocks. To do this we will utilise the 
Ising-chain domain wall representation introduced in  \cite{Chancellor19b}. That is for every position $\ell$ we add to the Hamiltonian 
\begin{equation}
\mathcal{H_\ell^\mathrm{(chain)}}= - \Lambda \left(\sum_{j=1}^{N-1} \sigma_{\ell N+j}^Z \sigma_{\ell N+j+1}^Z - \sigma^Z_{\ell N +1} + \sigma^Z_{\ell N + N}\right)~. \label{eq:H_chain}
\end{equation}
As shown in \cite{Chancellor19b}, taking $\Lambda$ to be much larger than every other energy scale in the overall Hamiltonian, these terms will constrain the system to remain in the ground subspace of the Hamiltonian, where exactly one spin position, $\alpha_\ell$ say, is frustrated for each $\ell$. 
These states are of the form 
\begin{equation}
\ket{11... 100 ...0}_\ell ~\implies~ \phi(\rho_\ell) \,=\, \phi_0 +  \alpha_\ell \xi ~,
\label{eq:gstate}
\end{equation}
where in the above the discretised field value is represented by the position $\alpha_\ell $ of the frustrated domain wall. Conversely the field value at the $\ell$'th position can be found by making the measurement
\begin{align}
\label{frust}
\phi(\rho_\ell) \,&=\,  \frac{1}{2} \sum_{j=1}^{N-1} (\phi_0+j\xi) \, \langle \sigma^Z_{\ell N + j+1 } -\sigma^Z_{\ell N + j} \rangle ~,
\end{align}
which only receives a contribution from frustrated spin position with $j=\alpha_\ell$. For later, it is useful to note that this is equivalent to 
\begin{align}
\label{frust2}
\phi(\rho_\ell) \,&=\,  \phi_0 + \frac{N\xi}{2} -  \frac{\xi}{2} \sum_{j=1}^{N} \, \langle \sigma^Z_{\ell N + j } \rangle ~.
\end{align}
In terms of $J_{ij}$ and $h_i$, adding the full set of Ising-chain Hamiltonians given by Eq.\eqref{eq:H_chain} corresponds to 
\begin{align} 
\label{eq:Jchain}
 J_{\ell N +i, m N + j}^{(\text{chain})} & =-\frac{\Lambda}{2}\,\delta_{\ell m} \,\otimes\,\left(\begin{array}{cccccc}
0 & 1\\
1 & 0 & 1\\
 & 1 & 0\\
 &  &  & \ddots\\
 &  &  &  & 0 & 1\\
 &  &  &  & 1 & 0
\end{array}\right)_{ij}  ~,
\end{align}
and an $h$ that is independent of $\ell$,
\begin{align} 
h^{(\text{chain})}_{\ell N + j}  & =\Lambda\,
(\delta_{j1}-\delta_{jN})~.
\label{eq:hchain}
\end{align}
This separates the system of spins into blocks of size $N$, each of which represents a field value. 

Moving on to $\mathcal{H}^{ (\mathrm QFT)}$, the potential is somewhat easier to deal with than the kinetic terms, because it can be encoded entirely in $h_i$. This is only to be expected because the $\phi_\ell$ are independent of each other in the potential which gives entirely localised contributions to the Hamiltonian. The value of $U(\phi(\rho_\ell))$ at each point follows directly from Eq.\eqref{frust}:
\begin{equation}
U(\phi(\rho_\ell)) \,=\, \frac{1}{2} \sum_{j=1}^{N-1} U(\phi_0+j\xi ) \, \langle \sigma^Z_{\ell N + j+1 } -\sigma^Z_{\ell N + j} \rangle  ~.
\end{equation}
This yields an additional contribution to the $h$ which is also independent of $\ell$: that is for all $\ell$ we have 
\begin{align}
h_{N \ell + j} ^{(\text{QFT})}\,&=\, 
\left\{ 
 \begin{array}{ll}
\vspace{0.2cm} 
{ \frac{\nu}{2} \left( U(\phi_{0}+(j-1) \xi)-U(\phi_{0}+j \xi) \right)}~;  & { j<N} \\
{ \frac{\nu}{2} U(\phi_{0}+(N-1) \xi) }~; &{ j=N }
\end{array}
\right.
\end{align}
It can also be convenient to write this in terms of $U$ derivatives
as 
\begin{align}
\label{eq:hqft}
h_{N \ell + j} ^{(\text{QFT})}\,&=\, 
\left\{ 
 \begin{array}{ll}
\vspace{0.2cm} { - \frac{\nu\xi }{2}  U'(\phi_{0}+j\xi) } ~; & { j<N} \\
{ \frac{\nu}{2} \left( U(\phi_{0}+(N-1) \xi) \right)}~;  &{ j=N }\, ,
\end{array}
\right.
\end{align}
which correctly gives $\phi(\rho_{\ell})$ of Eq.\eqref{frust2} in the event that we take
$U(\phi)=\phi$ (because we know that $\sigma_{\ell N}^{Z}=1$). Note
that in a system with arbitrary $c\neq0$, we would need to evaluate
$h^{(U)}\equiv\int d\rho\rho^{c}U$, so that $h_{\ell N+i}$ would acquire a prefactor of 
$(\ell\nu)^{c}.$ 

Up to this point the $M$-factors have been inert and there has been no
coupling between the fields at different positions in $\rho_{\ell}$.
At this stage the system would simply relax to $M$ decoupled
values of $\phi(\rho_\ell)$ that minimise $U$ in either one of its two vacua. This
changes once we include the derivatives in the kinetic terms, which contribute to the
bilinear interactions, $J$. These terms are discretised in $\rho$ as 
\begin{equation}
S_{KE}\equiv\int_{0}^{\Delta\rho}d\rho\frac{1}{2}\dot{\phi}^{2}=\lim_{M\rightarrow\infty}\sum_{\ell=1}^{M}\frac{1}{2\nu}\left(\phi(\rho_{\ell+1})-\phi(\rho_{\ell})\right)^{2},
\label{Eq:SKE}
\end{equation}
where $\nu=\Delta\rho/M$ scales so as to keep $\Delta\rho$ constant.
Inserting the discrete representation of the field values as well using Eq.\eqref{frust2},
we find 
\begin{eqnarray}
S_{KE} &= \sum_{\ell=1}^{M-1}\sum_{ij}^{N-1}\frac{\xi^{2}}{8\nu }\left[\sigma_{(\ell+1)N+i}^{Z}-\sigma_{\ell N+i}^{Z}\right]\times \qquad \qquad  \\
& \qquad\qquad\qquad\qquad\qquad \left[\sigma_{(\ell+1)N+j}^{Z}-\sigma_{\ell N+j}^{Z}\right].\nonumber
\end{eqnarray}
Hence the bilinear terms receive the additional
contribution: 
\begin{equation}
\medmath 
J_{\ell N +i, m N + j}^{(\text{QFT})}  =\frac{\xi^{2}}{8\nu}\,\left(\begin{array}{cccccc}
1 & -1\\
-1 & 2 & -1\\
 & -1 & 2 & -1\\
 &  &  & \ddots\\
 &  &  & -1 & 2 & -1\\
 &  &  &  & -1 & 1
\end{array}\right)_{\ell m}
\label{eq:Jqft}
\end{equation}
Now it is the $N\times N$ indices that are inert, because every $i$
couples to every $j$. 

Note that the diagonal parts of Eq.\ref{eq:Jqft} could be embedded in
the $h_{i}$ terms, using the fact that for valid single domain wall states we have  
$
\langle\sigma^Z_{\ell N + i} \sigma^Z_{\ell N + j} \rangle=\langle\sigma^Z_{\ell N + j}- \sigma^Z_{\ell N + i}  + 1 \rangle
$ for $j>i$. As bilinear terms may be hard to engineer on real devices, this may be desirable, but for the present study 
 it is more convenient to keep the kinetic terms entirely.  

Finally we must add terms to enforce a boundary condition. In the $c=0$ case it is sufficient to 
fix the endpoints of the solution in the two minima (so that, at the risk of confusion, the instanton solution 
itself approximates a physical domain wall). This can be done by adding a term ${\cal H}^{(BC)} = \frac{\Lambda'}{2} (\phi(0)+a)^2 +  \frac{\Lambda'}{2}   (\phi(\rho_M)-a)^2 $ with 
$\Lambda '$ being some other large parameter. This is simply an extra contribution to $h$ which follows directly from Eq.\eqref{eq:hqft}, of the form    
\begin{align}
\label{eq:hbc}
h_{N \ell + j} ^{(\text{BC})}\,&=\, 
\left\{ 
 \begin{array}{ll}
\vspace{0.2cm} 
-  {\Lambda'} (\phi_{0}+j \xi + a )~;  & { \ell = 1}, \forall j \\
-  {\Lambda'} (\phi_{0}+j \xi - a) ~; &{ \ell =M-1, \forall j }~.
\end{array}
\right.
\end{align}
Together with Eqs.(\ref{eq:Jchain},\ref{eq:hchain},\ref{eq:hqft},\ref{eq:Jqft}), this completes the encoding of the field theory
problem of Eq.\eqref{eq:bubble1}.  

\section{Implementation}

In Sec.~\ref{sec:encode} we have devised a method which encodes the problem of finding a solution to a quantum field theoretical problem, i.e. of finding a solution to Eq.~\ref{eq:bubble1}, into finding the ground state of the Hamiltonian of an Ising model. The latter can then be given an interpretation as the solution to Eq.~\ref{eq:bubble1} through Eq.~\ref{eq:gstate}, for each $\rho_l$ with $l \in [1,...,M]$. 
 To show that our approach is valid and converges to the correct solution $\phi(\rho)$, we now implement the method onto various annealing samplers, as provided by D-Wave \cite{D-Wave}. 

The quantum states are characterised by $NM$-tuples of the form $\left |11...100...0 \right>$ and the Hilbert space of the Ising model is therefore $2^{NM}$ dimensional. Sampling such a large vector space classically, with an exact sampler, while calculating the expectation value $\left<\mathcal{H} \right>$ for each state quickly becomes a computationally prohibitive task for $NM \gg 20$. Conversely, a discretisation with $NM \lesssim 20$ cannot give a reasonable approximation for the derivatives of Eq.~\ref{Eq:SKE}. 

Yet, not unlike protein-folding, in which a unique ground state is selected from an estimated number of $3^{300}$ so-called conformations within microseconds (known as Levinthal's paradox \cite{Levinthal}), a quantum annealer can in principle find a ground state of a Hamiltonian acting on a highly complex Hilbert space on a similar time scale, assuming there is a gap between the ground state and the other states of the system. 

While the next generation of annealing processors will have approximately 5,000 qubits, they will have limited connectivity
\cite{boothby}. Therefore in order to accommodate the more general Ising model required for our encoding, we  resorted to a hybrid asynchronous decomposition sampler (the Kerberos solver \cite{dwave-hybrid}), which can solve problems of arbitrary structure and size. 
To find the ground state efficiently, it applies in parallel classical tabu search algorithms, simulated annealing and D-Wave subproblem sampling on variables that have high-energy impact. Using this method we calculate the solution $\phi(\rho)$ to Eq.~\ref{eq:bubble1} for  $N=M=50$ in Fig.~\ref{fig:solutions}b.

\section{Conclusion}

Consequently, near-term applications of quantum devices can significantly enhance our ability to perform highly complex quantum field theoretical calculations. Focussing on the problem of calculating the classical instanton solution in Sec.~\ref{sec:problem}, we have proposed a method to formulate such problems as an Ising model in Sec.~\ref{sec:encode}, which we have solved on a D-Wave quantum annealer. Our method is highly flexible and can straightforwardly be generalised to encode any multi-dimensional differo-integral equation as an Ising model.

\bibliographystyle{mykp}
\bibliography{bibLibrary}  

\end{document}